\documentclass[aps,showpacs,twocolumn,superscriptaddress]{revtex4-1}
\usepackage{graphicx}
\usepackage{amsfonts}
\usepackage{amssymb}
\usepackage{amsmath}
\usepackage{natbib}
\usepackage{dcolumn}
\usepackage{bm}
\begin{document}
\title {Neutron star matter with strange interactions in a relativistic quark 
model}
\author{H .S. Sahoo}
\author{R. N. Mishra}
\author{D. K. Mohanty}
\affiliation{Department of Physics, Ravenshaw University, Cuttack-753 003, 
India}
\author{P.K. Panda}
\author{N. Barik}
\affiliation{Department of Physics, Utkal University, Bhubaneswar-751 004, 
India}
%\author{T. Frederico}
%\affiliation{ Instituto Tecn\'ologico de Aeron\'atica, DCTA,
%12228-900 S\~ao Jos\'e dos Campos, SP, Brazil}
%\date{\today}
\begin{abstract}
The effect of strange interactions in neutron star matter and the role of the 
strange meson-hyperon couplings are studied in a relativistic quark model 
where the confining 
interaction for quarks inside a baryon is represented by a phenomenological 
average potential in an equally mixed scalar-vector harmonic form. 
The hadron-hadron interaction in nuclear matter is then realized by 
introducing additional quark couplings to $\sigma$, $\omega$, $\rho$, 
$\sigma^*$ and $\phi$ mesons through mean-field approximations. The 
meson-baryon couplings are fixed through the SU(6) spin-flavor symmetry and 
the SU(3) flavor symmetry to determine the hadronic equation of state (EoS). 
We find that the SU(3) coupling set gives the potential depth 
between $\Lambda$s around $-5$ MeV and favours a stiffer EoS.The radius 
for the canonical neutron star lies within a range of $12.7$ to $13.1$ km. 

\end{abstract}

\maketitle
\section{Introduction}
The recent observation of gravitational waves from a binary neutron 
star coalescence by the Advanced LIGO and Virgo gravitational wave detectors, 
i.e., event GW170817 \cite{ligo} has provided new insight on the maximum mass 
as well as the radii distribution of neutron stars 
\cite{annala,janka,margalit,rezolla,shapiro}. Studies based on the GW170817 
observation put forth a stringent limit
on the radius corresponding to the $1.4 M_{\odot}$ mass neutron star, between
$9.9<R_{1.4}<13.6$ km. Such a limit sets a strong constraint on the equation 
of sate (EoS) of dense matter. The composition of dense matter relevant to 
neutron stars consists not only of nucleons and leptons but also several
exotic components such as hyperons, mesons as well as quark matter in
different forms and phases. Since at very high density it is energetically 
favourable for the creation of particles with strange content, it is expected 
that hyperons may appear in the inner core of neutron stars at densities 
$2-3$ times the normal saturation density $\rho_0=0.15$ fm$^{-3}$. The 
onset of this new degree of freedom softens the EoS and lowers the maximum 
mass. 

In the present work we study the properties of neutron stars, 
such as the mass, radius, and particle fractions by taking into 
consideration the effect of strangeness in neutron stars in a relativistic 
quark model, alternatively called the modified quark meson coupling model 
(MQMC) \cite{rnm,hss,hss2}. The MQMC model is based on  
confining relativistic independent quark potential model rather 
than a bag to describe the baryon structure in vacuum. In such a picture 
the quarks inside the baryon are considered to be independently confined 
by a phenomenological average potential with an equally mixed scalar-vector 
harmonic form. Such a potential has  characteristically simplifying features 
in converting the independent quark Dirac equation into a Schr\"odinger 
like equation for the upper component of Dirac spinor which can be solved 
easily. The implications of such potential forms in the Dirac framework 
has been studied earlier \cite{barik,prd}. 

In an earlier work \cite{hss2} we studied hyperon stars
in the MQMC model where the baryon-baryon interaction was realized through
$\sigma$, $\omega$, and $\rho$ mesons exchanges and the strange
quarks were considered as spectators. In the present attempt  we incorporate an
additional pair of hidden strange mesons $\sigma^*$ and $\phi$ 
\cite{pdg} which couple
only to the strange quark and the hyperons of the nuclear matter.
The relevant parameters of the interaction 
are obtained self-consistently by realizing the saturation properties 
such as binding energy and pressure. The hyperon couplings to the strange 
mesons are quite uncertain. 
%In the present work we have also 
%used an additional non-linear $\omega-\rho$ coupling to study its 
%effect on the radius of the star.
Fixing the coupling constants of hyperons with mesons to 
take into consideration the hyperon-hyperon inetraction is 
also a major difficulty. It is commom in literature to consider the hybrid 
SU(6) symmetry group \cite{pais} to fix the couplings of the hyperons with 
the vector mesons. We construct the EoS with three different coupling 
parameter sets. These are fixed using the SU(6) spin-flavor symmetry and  
by breaking the SU(6) symmetry of the 
isoscalar-vector meson to a more general SU(3) flavour symmetry group 
\cite{miyatsu} since such a scheme is expected to produce more massive neutron 
stars as wel as change their baryon composition. 
We also use the available hyperon-nucleon interaction
potential at saturation density for the
$\Lambda$, $\Sigma$ and $\Xi$ hyperons to $U_\Lambda = -28$ MeV,
$U_\Sigma = 30$ MeV and $U_\Xi = -18$ MeV respectively to determine the 
hyperon couplings to the vector $\omega$ meson. 

The paper is organized as follows: In Sec. II, a brief outline
of the model describing the baryon structure in vacuum is discussed and  
the baryon mass is then realized by appropriately taking into account the 
center-of-mass correction, pionic correction, and gluonic correction. 
The EOS is then developed in Sec. III. The results and discussions are 
made in Sec. IV. We summarize our findings in Sec. V.

\section{Modified quark meson coupling model}

The modified quark-meson coupling model has been broadly applied for the 
study of the bulk properties of symmetric and asymmetric nuclear matter. In an 
earlier work this model was used to study the role of hyperons in 
neutron stars without taking in to consideration the  
contribution of the hyperon-hyperon interactions \cite{rnm,hss,hss2}. We now 
extend this model to study the effect of strangeness in dense nuclear matter 
by including the contribution of the hidden strange mesons 
$\sigma^*$ and $\phi$. We begin by considering baryons as composed of 
three constituent quarks in a phenomenological flavor-independent confining 
potential, $U(r)$ in an equally mixed scalar and vector harmonic form inside 
the baryon \cite{rnm}, where 
\[
U(r)=\frac{1}{2}(1+\gamma^0)V(r),
\]
with 
\begin{equation}
V(r)=(ar^2+V_0),~~~~~ ~~~ a>0. 
\label{eq:1}
\end{equation}
Here $(a,~ V_0)$ are the potential parameters. The confining interaction  
provides the zeroth-order quark dynamics of the hadron.  In the medium, the 
quark field $\psi_q({\mathbf r})$ satisfies 
the Dirac equation
\begin{eqnarray}
&&[\gamma^0~(\epsilon_q-V_\omega-
\frac{1}{2} \tau_{3q}V_\rho-V_\phi)-{\vec \gamma}.{\vec p}\nonumber\\
&-&(m_q-V_\sigma-V_{\sigma^*})-U(r)]\psi_q(\vec r)=0
\end{eqnarray}
where $V_\sigma=g_\sigma^q\sigma_0$, $V_\omega=g_\omega^q\omega_0$, 
$V_\rho=g_\rho^q b_{03}$, $V_\phi=g_\phi^q\phi_0$ and 
$V_{\sigma^*}=g_{\sigma^*}^q\sigma_0^*$. Here $\sigma_0$, $\omega_0$, 
$b_{03}$, $\sigma_0^*$ and $\phi_0$ are the classical meson fields, and 
$g_\sigma^q$, $g_\omega^q$, $g_\rho^q$, $g_{\sigma_*}^q$ 
and $g_\phi^q$ are the quark couplings to  
the $\sigma$, $\omega$, $\rho$, $\sigma^*$ and $\phi$ mesons respectively. 
$m_q$ is the quark mass and $\tau_{3q}$ is the third component of the 
Pauli matrices. We can now define
\begin{equation}
\epsilon^{\prime}_q= (\epsilon_q^*-V_0/2)~~~ 
\mbox{and}~~~ m^{\prime}_q=(m_q^*+V_0/2),
\label{eprim}
\end{equation}
where the effective quark energy, 
$\epsilon_q^*=\epsilon_q-V_\omega-\frac{1}{2}\tau_{3q} V_\rho-V_\phi$ and 
effective quark mass, $m_q^*=m_q-V_\sigma-V_{\sigma^*}$. 
We now introduce $\lambda_q$ 
and $r_{0q}$ as
\begin{equation}
(\epsilon^{\prime}_q+m^{\prime}_q)=\lambda_q~~
~~\mbox{and}~~~~r_{0q}=(a\lambda_q)^{-\frac{1}{4}}.
\label{eq:8}
\end{equation}

The ground-state quark energy can be obtained from the eigenvalue condition
\begin{equation}
(\epsilon^{\prime}_q-m^{\prime}_q)\sqrt \frac{\lambda_q}{a}=3.
\label{eq:11}
\end{equation}
The solution of equation \eqref{eq:11} for the quark  energy
$\epsilon^*_q$ immediately leads to
the mass of baryon in the medium in zeroth order as
\begin{equation}
E_B^{*0}=\sum_q~\epsilon^*_q
\label{eq:12}
\end{equation}

%\section{Effective mass of baryon}
We next consider the spurious center-of-mass 
correction $\epsilon_{c.m.}$, the pionic correction $\delta M_{B}^\pi$ 
for restoration of chiral symmetry, and the short-distance one-gluon 
exchange contribution $(\Delta E_B)_g$ to the zeroth-order baryon 
mass in the medium. 

Here, we extract the center of mass energy to first order in the 
difference between the fixed center and relative quark co-ordinate, 
using the method described by Guichon {\it et al.} \cite{guichon}. 
The centre of mass correction is given by:
\begin{equation}
e_{c.m.}=e_{c.m.}^{(1)}+e_{c.m.}^{(2)},
\end{equation}
where,

\begin{equation}
e_{c.m.}^{(1)}=\sum_{i=1}^3{\left[\frac{m_{q_i}}{\sum_{k=1}^3 m_{q_k}}\frac{6}
{r_{0q_i}^2(3\epsilon'_{q_i}+m'_{q_i})}\right]}
\end{equation}

\begin{widetext}
\begin{eqnarray}
e_{c.m.}^{(2)}&=&\frac{a}{2}\bigg[\frac{2}{\sum_k m_{q_k}}
\sum_im_i\langle r_i^2\rangle
+\frac{2}{\sum_k m_{q_k}}\sum_im_i\langle \gamma^0(i)r_i^2\rangle
-\frac{3}{(\sum_k m_{q_k})^2}\sum_im_i^2\langle r_i^2\rangle\nonumber\\
&-&\frac{1}{(\sum_k m_{q_k})^2}\sum_i\langle \gamma^0(1)m_i^2r_i^2\rangle-
\frac{1}{(\sum_k m_{q_k})^2}\sum_i\langle \gamma^0(2)m_i^2r_i^2\rangle-
\frac{1}{(\sum_k m_{q_k})^2}\sum_i\langle \gamma^0(3)m_i^2r_i^2\rangle\bigg]
\end{eqnarray}
\end{widetext}

 In the above, we have used for $i=(u,d,s)~~~$ and $k=(u,d,s)$ and the 
various quantities are defined as

\begin{equation}
\langle r_i^2\rangle = \frac{(11\epsilon_{qi}'+ m_{qi}')r^2_{0qi}}
{2(3\epsilon_{qi}'+ m_{qi}')}
\end{equation}

\begin{equation}
\langle\gamma^0(i)r_i^2\rangle=\frac{(\epsilon_{qi}'+ 11 m_{qi}')r^2_{0qi}}
{2(3\epsilon_{qi}'+ m_{qi}')}
\end{equation}

\begin{equation}
\langle\gamma^0(i)r_j^2\rangle_{i\neq j}=\frac{(\epsilon_{qi}'+ 3 m_{qi}')
\langle r^2_j\rangle}{3\epsilon_{qi}'+ m_{qi}'}
\end{equation}

The pseudo-vector nucleon pion coupling constant, $f_{NN\pi}$ can be obtained 
from Goldberg-Treiman relations by using the axial-vector coupling constant 
value $g_A$ in the model as
\begin{equation}
{\sqrt{4\pi}}\frac{f_{NN\pi}}{m_\pi}=\frac{g_A(N)}{2f_\pi},
\end{equation}
where
\begin{equation}
g_A(n\rightarrow p)= \frac{5}{9}\frac{(5\epsilon_u^{\prime}
+7m_u^{\prime})} {(3\epsilon_u^{\prime}+m_u^{\prime})}.
\end{equation}
The pionic corrections in the model for the nucleons become
\begin{equation}
\delta M_{N}^\pi=- \frac{171}{25}I_{\pi}f_{NN\pi}^2.
\end{equation}
Taking $w_k=(k^2+m_\pi^2)^{1/2}$
$I_{\pi}$ becomes
\begin{equation}
I_{\pi}=\frac{1}{\pi{m_{\pi}}^2}\int_{0}^{\infty}dk. 
\frac{k^4u^2(k)}{w_k^2},
\end{equation}
with the axial vector nucleon form factor given as
\begin{equation}
u(k)=\Big[1-\frac{3}{2} \frac{k^2}{{\lambda}_q(5\epsilon_q^{\prime}+
7m_q^{\prime})}\Big]e^{-k^2r_0^2/4} \ .
\end{equation}
The pionic correction for $\Sigma^{0}$ and $\Lambda^{0}$ become
\begin{equation}
\delta M_{\Sigma^{0}}^{\pi}=-{\frac{12}{5}}f_{NN\pi}^2I_{\pi},
\end{equation}
\begin{equation}
\delta M_{\Lambda^{0}}^{\pi}=-{\frac{108}{25}}f_{NN\pi}^2I_{\pi}.
\end{equation}
Similarly the pionic correction for $\Sigma^{-}$ and $\Sigma^{+}$ is
\begin{equation}
\delta M_{\Sigma^{+},\Sigma^{-}}^{\pi}=-{\frac{12}{5}}f_{NN\pi}^2I_{\pi}.
\end{equation}
The pionic correction for $\Xi^{0}$ and $\Xi^{-}$ is
\begin{equation}
\delta M_{\Xi^{-},\Xi^{0}}^{\pi}=-{\frac{27}{25}}f_{NN\pi}^2I_{\pi}.
\end{equation}

The one-gluon exchange interaction is provided by the interaction Lagrangian
density
\begin{equation}
{\cal L}_I^g=\sum J^{\mu a}_i(x)A_\mu^a(x) \ ,
\end{equation}
where $A_\mu^a(x)$ are the octet gluon vector-fields and $J^{\mu a}_i(x)$ is 
the $i$-th quark color current. The gluonic correction can be separated in two
pieces, namely, one from the color electric field ($E^a_i$) and another 
from the magnetic field ($B^a_i$)
generated by the $i$-th quark color current density
\begin{equation}
J^{\mu a}_i(x)=g_c\bar\psi_q(x)\gamma^\mu\lambda_i^a\psi_q(x) \ ,
\end{equation}
with $\lambda_i^a$ being the usual Gell-Mann $SU(3)$ matrices and
$\alpha_c=g_c^2/4\pi$. The contribution to the mass 
can be written as a sum of color electric and color magnetic part as
\begin{equation}
(\Delta E_B)_g=(\Delta E_B)_g^E+(\Delta E_B)_g^M \ ,
\end{equation}
where
\begin{eqnarray}
(\Delta E_B)_g^E &=&\frac{1}{8\pi}\sum_{i,j}\sum_{a=1}^8\int\frac{d^3r_id^3r_j}
{|r_i-r_j|}\nonumber\\
&\times&\langle B |J^{0 a}_i(r_i)J^{0 a}_j(r_j)|B\rangle \ ,
\end{eqnarray}
and
\begin{eqnarray}
(\Delta E_B)_g^M&=&-\frac{1}{8\pi}\sum_{i,j}\sum_{a=1}^8\int\frac{d^3r_id^3r_j}
{|r_i-r_j|}\nonumber\\
&\times& \langle B |\vec J^a_i(r_i)\vec J^a_j(r_j)|B\rangle \ .\\ \nonumber
\end{eqnarray}

Finally, taking into account the specific quark flavor and spin configurations
in the ground state baryons and using the relations
$\langle\sum_a(\lambda_i^a)^2\rangle =16/3$ and
$\langle\sum_a(\lambda_i^a\lambda_j^a)\rangle_{i\ne j}=-8/3$ for
baryons, one can write the energy correction due to
color electric contribution, as
\begin{equation}
(\Delta E_B)_g^E={\alpha_c}(b_{uu}I_{uu}^E+b_{us}I_{us}^E+b_{ss}I_{ss}^E) \ ,   
\label{enge}
\end{equation}
and due to color magnetic contributions, as
\begin{equation}
(\Delta E_B)_g^M={\alpha_c}(a_{uu}I_{uu}^M+a_{us}I_{us}^M+a_{ss}I_{ss}^M) \ ,  
\label{engm}
\end{equation}
where $a_{ij}$ and $b_{ij}$ are the numerical coefficients depending on each
baryon and are given in Table \ref{table0}. In the above, we have

\begin{table}[t]
\renewcommand{\arraystretch}{1.4}
\setlength\tabcolsep{3pt}
\begin{tabular}{|c|c|c|c|c|c|c|}
\hline
Baryon     & $a_{uu}$ & $a_{us}$ & $a_{ss}$ & $b_{uu}$ & $b_{us}$ & $b_{ss}$\\
\hline
$N$        & -3  &  0  & 0 & 0 &  0 & 0\\
$\Lambda$  & -3  &  0  & 0 & 1 & -2 & 1\\
$\Sigma$   &  1  & -4  & 0 & 1 & -2 & 1\\
$\Xi$      &  0  & -4  & 1 & 1 & -2 & 1\\
\hline
\end{tabular}
\caption{\label{table0}The coefficients $a_{ij}$ and $b_{ij}$ used in the 
calculation of the color-electric and and color-magnetic energy 
contributions due to one-gluon exchange.}
\end{table}

\begin{eqnarray}
I_{ij}^{E}=\frac{16}{3{\sqrt \pi}}\frac{1}{R_{ij}}\Bigl[1-
\frac{\alpha_i+\alpha_j}{R_{ij}^2}+\frac{3\alpha_i\alpha_j}{R_{ij}^4}
\Bigl]
\nonumber\\
I_{ij}^{M}=\frac{256}{9{\sqrt \pi}}\frac{1}{R_{ij}^3}\frac{1}{(3\epsilon_i^{'}
+m_{i}^{'})}\frac{1}{(3\epsilon_j^{'}+m_{j}^{'})} \ ,
\end{eqnarray}
where
\begin{eqnarray}
R_{ij}^{2}&=&3\Bigl[\frac{1}{({\epsilon_i^{'}}^2-{m_i^{'}}^2)}+
\frac{1}{({\epsilon_j^{'}}^2-{m_j^{'}}^2)}\Bigl]
\nonumber\\
\alpha_i&=&\frac{1}{ (\epsilon_i^{'}+m_i^{'})(3\epsilon_i^{'}+m_{i}^{'})} \ .
\end{eqnarray}
The color electric contributions to the bare mass for nucleon 
$(\Delta E_N)_g^{E} = 0$. Therefore the one-gluon contribution for nucleon 
becomes
\begin{equation}
(\Delta E_N)_g^{M}=-{\frac{256\alpha_c}{3\sqrt{\pi}}}
\Big [{\frac{1}{(3{\epsilon}_u^{\prime}+m_u^{\prime})^2R_{uu}^3}}\Big ]\\
\end{equation}

The one-gluon contribution for $\Sigma^{+} ,\Sigma^{-}$ becomes
\begin{eqnarray}
(\Delta E_{\Sigma^{+} ,\Sigma^{-}})_g^{E}&=&{\alpha_c}{\frac{16}{3\sqrt{\pi}}}
\Bigg [{\frac{1}{R_{uu}}}\left(1-{\frac{2\alpha_u}{R_{uu}^2}}-
{\frac{3\alpha_u^2}{R_{uu}^4}}\right) \nonumber \\   
&-&{\frac{2}{R_{us}}}\left(1-{\frac{\alpha_u+\alpha_s}{R_{us}^2}} +
{\frac{3\alpha_u\alpha_s}{R_{us}^4}}\right) \nonumber\\
&+&{\frac{1}{R_{ss}}}\left(1-{\frac{2\alpha_s}{R_{ss}^2}}+
{\frac{3\alpha_s^2}{R_{ss}^4}}\right)\Bigg ]
\end{eqnarray}

\begin{eqnarray}
(\Delta E_{\Sigma^{+} ,\Sigma^{-}})_g^{M}&=&{\frac{256\alpha_c}{9\sqrt{\pi}}}
\Bigg [{\frac{1}{(3{\epsilon}_u^{\prime}+m_u^{\prime})^2R_{uu}^3}}\nonumber\\ 
&-&{\frac{4}
{R_{us}^3(3{\epsilon}_u^{\prime}+m_u^{\prime})(3{\epsilon}_s^{\prime}+
m_s^{\prime})}}\Bigg ]
\end{eqnarray}
\begin{equation}
(\Delta E_{\Sigma^{+} ,\Sigma^{-}})_g=(\Delta E_{\Sigma^{+} ,
\Sigma^{-}})_g^{E}+(\Delta E_{\Sigma^{+} ,\Sigma^{-}})_g^M
\end{equation}
The gluonic correction for $\Sigma^{0}$ is
\begin{eqnarray}
(\Delta E_{\Sigma^{0}})_g^{E}&=&{\alpha_c}{\frac{16}{3\sqrt{\pi}}}
\Bigg [{\frac{1}{R_{uu}}}\left(1-{\frac{2\alpha_u}{R_{uu}^2}}-
{\frac{3\alpha_u^2}{R_{uu}^4}}\right)  \nonumber \\   
&-&{\frac{2}{R_{us}}}\left(1-{\frac{\alpha_u+\alpha_s}{R_{us}^2}} +
{\frac{3\alpha_u\alpha_s}{R_{us}^4}}\right) \nonumber\\
&+&{\frac{1}{R_{ss}}}\left(1-{\frac{2\alpha_s}{R_{ss}^2}}+
{\frac{3\alpha_s^2}{R_{ss}^4}}\right)\Bigg ]
\end{eqnarray}

\begin{eqnarray}
(\Delta E_{\Sigma^{0}})_g^{M}&=&{\frac{256\alpha_c}{9\sqrt{\pi}}}
\Bigg [{\frac{1}{(3{\epsilon}_u^{\prime} +m_u^{\prime})^2R_{uu}^3}}\nonumber\\ 
&-&{\frac{4}
{R_{us}^3(3{\epsilon}_u^{\prime}+m_u^{\prime})(3{\epsilon}_s^{\prime}+
m_s^{\prime})}}\Bigg ]
\end{eqnarray}
\begin{equation}
(\Delta E_{\Sigma^{0}})_g=(\Delta E_{\Sigma^{0}})_g^{E}
+(\Delta E_{\Sigma^{0}})_g^M
\end{equation}

The gluonic correction for $\Lambda$ is
\begin{equation}
(\Delta E_{\Sigma^{0}})_g^{E}=(\Delta E_{\Lambda})_g^{E}
\end{equation}
The color magnetic contribution is different
\begin{equation}
(\Delta E_{\Lambda})_g^{M}=-{\frac{256\alpha_c}{3\sqrt{\pi}}}\Bigg [
{\frac{1}{(3{\epsilon}_u^{\prime} +m_u^{\prime})^2R_{uu}^3}}\Bigg ]
\end{equation}
\begin{equation}
(\Delta E_{\Lambda})_g=(\Delta E_{\Lambda})_g^{E}
+(\Delta E_{\Lambda})_g^M
\end{equation}

The color electric contributions for $\Xi^{-}$ and $\Xi^{0}$ are same as
that of $\Sigma^{0}$ or $\Lambda^{0}$ but the color magnetic contributions
to the correction of masses of baryon  are different:
\begin{eqnarray}
(\Delta E_{\Xi^{-} ,\Xi^{0}})_g^{M}&=&{\frac{256\alpha_c}{9\sqrt{\pi}}}
\Bigg [{\frac{1}{(3{\epsilon}_s^{\prime}+m_s^{\prime})^2R_{ss}^3}}\nonumber\\ 
&-&{\frac{4}
{R_{us}^3(3{\epsilon}_u^{\prime}+m_u^{\prime})(3{\epsilon}_s^{\prime}+
m_s^{\prime})}}\Bigg ]
\end{eqnarray}

Finally, the gluonic correction for $\Xi^{-}$ and $\Xi^{0}$ is given by:
\begin{equation}
(\Delta E_{\Xi^{-} ,\Xi^{0}})_g=(\Delta E_{\Xi^{-} ,\Xi^{0}})_g^{E}+
(\Delta E_{\Xi^{-} ,\Xi^{0}})_g^M
\end{equation}

Treating all energy corrections independently, the 
mass of the baryon in the medium becomes 
\begin{equation}
M_B^*=E_B^{*0}-\epsilon_{c.m.}+\delta M_B^\pi+(\Delta E_B)^E_g+
(\Delta E_B)^M_g.
\label{mass}
\end{equation}

\section{The Equation of state}
The total energy density and pressure at a 
particular baryon density, encompassing all the members of the baryon octet, 
for the nuclear matter in $\beta$-equilibrium can be found as
\begin{eqnarray}
{\varepsilon}&=&\frac{1}{2}m_\sigma^2 \sigma_0^2+\frac{1}{2}m_{\sigma^*}^2 
\sigma_0^{*2}+
\frac{1}{2}m_\omega^2 \omega^2_0+\frac{1}{2}m_\phi^2 \phi^2_0
\nonumber\\ &+&
\frac{1}{2}m_\rho^2 \rho^2_{03}+\frac{\gamma}{2\pi^2}\sum_{B}\int_0^{k_B} k^2dk \sqrt{k^2+{M_B^*}^2}\nonumber\\
&-& g_\omega^2g_\rho^2\Lambda_v\rho_{03}^2\omega_0^2\nonumber\\
&+&\sum_{l}\frac{1}{\pi^2}\int_0^{k_l}k^2dk[k^2+m_l^2]^{1/2}
\end{eqnarray}

\begin{eqnarray}
P&=&-~\frac{1}{2}m_\sigma^2 \sigma_0^2-~\frac{1}{2}m_{\sigma^*}^2 \sigma_0^{*2}
+\frac{1}{2}m_\omega^2 \omega^2_0
\nonumber\\
&+&\frac{1}{2}m_\phi^2 \phi^2_0 + \frac{1}{2}m_\rho^2 \rho_{03}^2 
+g_\omega^2g_\rho^2\Lambda_v\rho_{03}^2\omega_0^2\nonumber\\ 
&+&\frac{\gamma}{6\pi^2}\sum_{B}\int_0^{k_B} \frac{k^4~dk}
{\sqrt{k^2+{M_B^*}^2}} \nonumber\\
&+& \frac{1}{3}\sum_{l}\frac{1}{\pi^2}\int_0^{k_l}
\frac{k^4dk}{[k^2+m_l^2]^{1/2}}
\end{eqnarray}
%
%\begin{subequations}
%\begin{eqnarray}
%\label{engd}
%{\cal E} &=&\frac{1}{2}m_\sigma^2 \sigma_0^2+\frac{1}{2}m_\omega^2 \omega^2_0
%+\frac{1}{2}m_\rho^2 b^2_{03} + 3g_\omega^2g_\rho^2\Lambda_\nu b_{03}^2\omega_0^2
%\nonumber\\ &+&\frac{\gamma}{2\pi^2}\sum_{B}\int ^{k_{f,B}} 
%[k^2+{M_B^*}^2]^{1/2}k^2dk
%\nonumber\\
%&+&\sum_{l}\frac{1}{\pi^2}\int_0^{k_l}[k^2+m_l^2]^{1/2}k^2dk,\\
%P&=&-~\frac{1}{2}m_\sigma^2 \sigma_0^2+\frac{1}{2}m_\omega^2 \omega^2_0+
%\frac{1}{2}m_\rho^2 b_{03}^2+ g_\omega^2g_\rho^2\Lambda_\nu b_{03}^2\omega_0^2
%\nonumber\\
%&+&\frac{\gamma}{6\pi^2}\sum_{B}\int ^{k_{f,B}} \frac{k^4~ dk}
%{[k^2+{M_B^*}^2]^{1/2}} \nonumber\\
%&+& \frac{1}{3}\sum_{l}\frac{1}{\pi^2}\int_0^{k_l}
%\frac{k^4dk}{[k^2+m_l^2]^{1/2}},
%\end{eqnarray} 
%\end{subequations}
where $\gamma=2$ is the spin degeneracy factor for nuclear matter,
$B=N,\Lambda,~\Sigma^{\pm},~\Sigma^0,~\Xi^-,~\Xi^0$ and $l=e,\mu$.
In the above expression for the energy density and pressure, a 
nonlinear $\omega-\rho$ coupling term is introduced with coupling
coefficient, $\Lambda_\nu$ \cite{horowitz01}.

%Another important quantity for the study of nuclear matter is the symmetry 
%energy, which is defined as
%\begin{equation}
%{\cal E}_{sym}(\rho_B)=\frac{k^2}{6E_N^{*2}}
%+\frac{g_\rho^2} {8m_\rho^{2}}\rho_B
%\label{engs}
%\end{equation}
%where $E_N^*=\sqrt{k^2+M_N^{*2}}$, the index $N=n,p$ for neutrons and protons. 
%The slope of the 
%symmetry energy $L$ is then obtained as,
%\begin{equation}
%L=3\rho_0\dfrac{\partial {\cal E}_{sym}(\rho_B)}{\partial \rho_B}
%\Bigg|_{\rho_B=\rho_0}
%\end{equation}
For obtaining a constraint on the quark mass we use the value of 
compressibility given by,
\begin{equation}
K=9\left[\frac{dP}{d\rho_B}\right]_{\rho_B=\rho_0}
\end{equation}

The chemical potentials, necessary to define the $\beta-$ 
equilibrium conditions, are given by
\begin{equation}
\mu_B=\sqrt{k_B^2+{M_B^*}^2}+g_\omega\omega_0+g_\rho\tau_{3B}b_{03}+
g_\phi\phi_0
\end{equation}
where $\tau_{3B}$ is the isopsin projection of the baryon B.

The lepton Fermi momenta are the positive real solutions of
$(k_e^2 + m_e^2)^{1/2} =  \mu_e$ and
$(k_\mu^2 + m_\mu^2)^{1/2} = \mu_\mu$. The equilibrium composition
of the star is obtained by solving the equations of motion of meson fields in 
conjunction with the charge neutrality condition, given in equation 
(\ref{neutral}),  
at a given total baryonic density $\rho = \sum_B \gamma k_B^3/(6\pi^2)$. 
The effective masses of the baryons are
obtained self-consistently in this model.

Since we consider the octet baryons, the presence of strange baryons 
in the matter plays a significant role. We define the strangeness 
fraction as 
\begin{equation}
f_s=\frac{1}{3}\frac{\sum_i |s_i|\rho_i}{\rho}.
\end{equation}
Here $s_i$ refers to the strangeness number of baryon $i$ and  
$\rho_i$ is defined as $\rho_i=\gamma k_{Bi}^3/(6\pi^2)$.

For stars in which the strongly interacting particles are baryons, the
composition is determined by the requirements of charge neutrality
and $\beta$-equilibrium conditions under the weak processes
$B_1 \to B_2 + l + {\overline \nu}_l$ and $B_2 + l \to B_1 + \nu_l$.
After deleptonization, the charge neutrality condition yields
\begin{equation}
q_{\rm tot} = \sum_B q_B \frac{\gamma k_B^3}{6\pi^2}
+ \sum_{l=e,\mu} q_l \frac{k_l^3}{3\pi^2}  = 0 ~,
\label{neutral}
\end{equation}
where $q_B$ corresponds to the electric charge of baryon species $B$
and $q_l$ corresponds to the electric charge of lepton species $l$. Since
the time scale of a star is effectively infinite compared to the weak
interaction time scale, weak interaction violates strangeness conservation.
The strangeness quantum number is therefore not conserved
in a star and the net strangeness is determined by the condition of
$\beta$-equilibrium which for baryon $B$ is then given by
$\mu_B = b_B\mu_n - q_B\mu_e$, where $\mu_B$ is the chemical potential
of baryon $B$ and $b_B$ its baryon number. Thus the chemical potential of any
baryon can be obtained from the two independent chemical potentials $\mu_n$
and $\mu_e$ of neutron and electron respectively. 
%The hyperon couplings to the 
%mesons are fixed using two different approaches through symmetry 
%considerations under SU(6) symmetry and SU(3) symmetry.
\subsection{Baryon-Meson coupling constants}
The Baryon-Meson coupling coupling constants are given by, 
$$g_{\sigma B}=x_{\sigma B}~ g_{\sigma N},~~g_{\omega B}=x_{\omega B}~ 
g_{\omega N}, ~~g_{\rho B}=x_{\rho B}~ g_{\rho N},$$
where $x_{\sigma B}$, $x_{\omega B}$ and $x_{\rho B}$ are equal to $1$ for the
nucleons and acquire different values in different parameterisations for the
other baryons. We note that the $s$-quark is unaffected by the $\sigma$- and 
$\omega$ mesons i.e. $g_\sigma^s=g_\omega^s=0$.  
To take into account the effect of the strange quark, we include the 
strange mesons $\sigma^*$ and $\phi$ with couplings $g_{\sigma^*}$ and 
$g_\phi$ respectively. The iso-scalar scalar and iso-scalar vector 
couplings for the nucleons $g_\sigma^q$ and $g_\omega$ are fitted to the 
saturation density and 
binding energy for nuclear matter. The iso-vector vector coupling to the 
nucleon $g_\rho$ is set by fixing the symmetry energy at $J=32.0$ MeV. 
The hyperon coupling ratios are determined after the coupling constants 
$g_\sigma^q$, $g_\omega$ and $g_\rho$ for the nucleon sector are determined. 
 
The 
vector meson couplings to the hyperons are fixed using three different 
approaches to get three parameter sets, Set I, Set II and Set III. 
In the first set we use the SU(6) 
spin-flavor symmetry \cite{cdover,agal} as follows,
\begin{eqnarray}  
&&\frac{1}{3}g_{\omega N}=\frac{1}{2}g_{\omega\Lambda}=
\frac{1}{2}g_{\omega\Sigma}=g_{\omega\Xi},\nonumber\\
&&g_{\rho N}=g_{\rho\Lambda}=\frac{1}{2}g_{\rho\Sigma}=g_{\rho\Xi},\\
&&2g_{\phi\Lambda}=2g_{\phi\Sigma}=g_{\phi\Xi}=
-\frac{2\sqrt{2}}{3}g_{\omega N},\nonumber
\end{eqnarray}
with $g_{\phi N}=0$. 

In Set II, we adjust the $\omega$-hyperon coupling strengths 
($x_{\omega\Lambda}$, $x_{\omega\Sigma}$ and $x_{\omega\Xi}$) to the 
hyperon-nucleon interaction potential at saturation density for the 
$\Lambda$, $\Sigma$ and $\Xi$ hyperons with $U_\Lambda = -28$ MeV,
$U_\Sigma = 30$ MeV and $U_\Xi = -18$ MeV respectively
using the relation 
$U_B=-(M_B-M_B^*)+x_{{\omega}B}g_{\omega}\omega_0$ for $B={\Lambda},
{\Sigma}$ and $\Xi$. 
We keep the $\rho$-coupling to the hyperons same as that of the nucleons and 
set $g_{\phi N}=0$.    

For Set III, we extend the SU(6) spin-flavor symmetry to the SU(3) flavor 
symmetry. The SU(3) group with three flavors of quarks (up, down, strange) is 
regarded as the symmetry group for strong interactions. 
We follow the scheme given in \cite{miyatsu,nijmegen,swart,rijken,weiss} 
using the matrix 
representations for the baryon octet and meson nonet (singlet state and 
octet state). In such a scheme, the $\omega$ and $\phi$ mesons are 
described in terms of the pure singlet, $|1\rangle$, and octet, $|8\rangle$, 
states as 
\begin{eqnarray}  
\omega=\cos\theta_\nu|1\rangle+\sin\theta_\nu|8\rangle \nonumber\\
\phi=-\sin\theta_\nu|1\rangle+\cos\theta_\nu|8\rangle,
\end{eqnarray}
with $\theta_\nu$ being the mixing angle.  
The mixing angle is fixed using the 
Nijmegen extended-soft-core (ESC) model \cite{nijmegen}. The values of the 
$\theta_\nu$ and the $z$ ($z\equiv g_8/g_1$, coupling ratio of the octet to 
singlet coupling constants), suggested in the ESC model are given as
\begin{equation}   
\theta_\nu=37.50^o,~~~~~ z=0.1949.
\end{equation}
Then the relations of the vector meson-hyperon coupling constants in SU(3) 
are given as,
\begin{eqnarray}
&g_{\omega\Lambda}&=g_{\omega\Sigma}=
\frac{1}{1+\sqrt{3}z\tan\theta_\nu}g_{\omega N},\nonumber\\
&g_{\omega\Xi}&=\frac{1-\sqrt{3}z\tan\theta_\nu}{1+\sqrt{3}z\tan\theta_\nu}
g_{\omega N},\\
&g_{\phi N}&=
\frac{\sqrt{3}z-\tan\theta_\nu}{1+\sqrt{3}z\tan\theta_\nu}g_{\omega N},
\nonumber\\
&g_{\phi\Lambda}&=g_{\phi\Sigma}=
\frac{-\tan\theta_\nu}{1+\sqrt{3}z\tan\theta_\nu}g_{\omega N},\nonumber\\
&g_{\phi\Xi}&=-\frac{\sqrt{3}z+\tan\theta_\nu}{1+\sqrt{3}z\tan\theta_\nu}
g_{\omega N}.\nonumber\\
\label{su3coupling}
\end{eqnarray}

We note here that the couplings of the hyperons to the $\sigma$-meson need 
not be fixed since we determine the effective masses of the hyperons 
self-consistently. In all the three sets considered above we have 
fixed $g_\sigma^*=2$ considering a weak hyperon-hyperon coupling.

The vector mean-fields $\omega_0$, $b_{03}$ and $\phi_0$ are determined through
\begin{eqnarray}
&&\omega_0=\frac{g_\omega}{{m_\omega^*}^2} \sum_B x_{\omega B}\rho_B,~~~~
b_{03}=\frac{g_\rho}{{2m_\rho^*}^2} \sum_B x_{\rho B}\tau_{3B}\rho_B,\nonumber\\
&&\phi_0=\frac{g_\phi}{{m_\phi^*}^2} \sum_B x_{\phi B}\rho_B
\label{omg}
\end{eqnarray}
where ${m_\omega^*}^2=m_\omega^2+2\Lambda_{\nu}g_\rho^2g_\omega^2b_{03}^2$, 
${m_\rho^*}^2=m_\rho^2+2\Lambda_{\nu}g_\rho^2g_\omega^2\omega_0^2$, 
$g_\omega=3 g_\omega^q$ and $g_\rho= g_\rho^q$.
Finally, the scalar mean-fields $\sigma_0$ and $\sigma_0^*$ are fixed by
\begin{equation}
\frac{\partial {\cal E }}{\partial \sigma_0}=0,~~~
\frac{\partial {\cal E }}{\partial \sigma_0^*}=0
\label{sig}
\end{equation}

Using the EOS determined above for the three parameter sets, 
we can obtain the relation between the mass and 
radius of a star with its central density by integrating 
the Tolman-Oppenheimer-Volkoff (TOV) equations \cite{tov} given by,
\begin{equation}
\frac{dP}{dr}=-\frac{G}{r}\frac{\left[{\cal E}+P\right ]\left[M+
4\pi r^3 P\right ]}{(r-2 GM)},
\label{tov1}
\end{equation}
\begin{equation}
\frac{dM}{dr}= 4\pi r^2 {\cal E} ,
\label{tov2}
\end{equation}
with $G$ as the gravitational constant and $M(r)$ as the enclosed gravitational
mass. We have used $c=1$. Given an EOS, these equations can be integrated 
from the origin as an initial value problem for a given choice of the 
central energy density, $(\varepsilon_0)$.
Of particular importance is the maximum mass obtained from and the 
solution of the TOV equations. 
The value of $r~(=R)$, where the pressure vanishes defines the
surface of the star. 

\section{Results and Discussion}
There are two potential parameters in the relativistic quark model, 
$a$ and  $V_0$ which are obtained by fitting the nucleon mass $M_N=939$ MeV 
and charge radius of the proton $\langle r_N\rangle=0.84$ fm in free space. 
Keeping the value of the potential parameter $a$ same as that for nucleons, 
we obtain $V_0$ for the $\Lambda$, $\Sigma$ and $\Xi$ baryons by fitting 
their respective masses to $M_{\Lambda}=1115.6$ MeV, $M_{\Sigma}=1193.1$ MeV 
and $M_{\Xi}=1321.3$ MeV. The mass of the $u$,$d$ quarks is fixed at $200$ MeV 
and the mass of the $s$ quark is fixed at $300$ MeV.
The set of potential parameters for the baryons are given in 
Table \ref{table1}. 
 
\begin{table}[b]
\renewcommand{\arraystretch}{1.4}
\setlength\tabcolsep{3pt}
\begin{tabular}{ccc}
\hline
\hline
Baryon     & $M_B$(MeV)&$V_0$(MeV)\\
\hline
$N$        & 939    &  5.44  \\
$\Lambda$  & 1115.6 &  28.00  \\
$\Sigma$   & 1193.1 &  43.29  \\
$\Xi$      & 1321.3 &  54.17 \\
\hline
\hline
\end{tabular}
\caption{\label{table1}The potential parameter $V_0$ for different baryons 
obtained for the
quark mass $m_u=m_d=200$ MeV, $m_s=300$ MeV with $a=0.722970$~fm$^{-3}$.}
\end{table}

The incompressibility $K$ of symmetric nuclear matter for quark mass 
$200$ MeV comes out to be $242.41$ MeV. 
Recent measurements \cite{umesh} extracted from doubly-magic nuclei 
like $^{208}$Pb constrain the value of $K$ to be around $240\pm20$.
%%%%%%%%%%%%%%%%
%%%%%%%%%%%%%%%%%%%%%
\begin{figure}
\includegraphics[width=8.cm,angle=0]{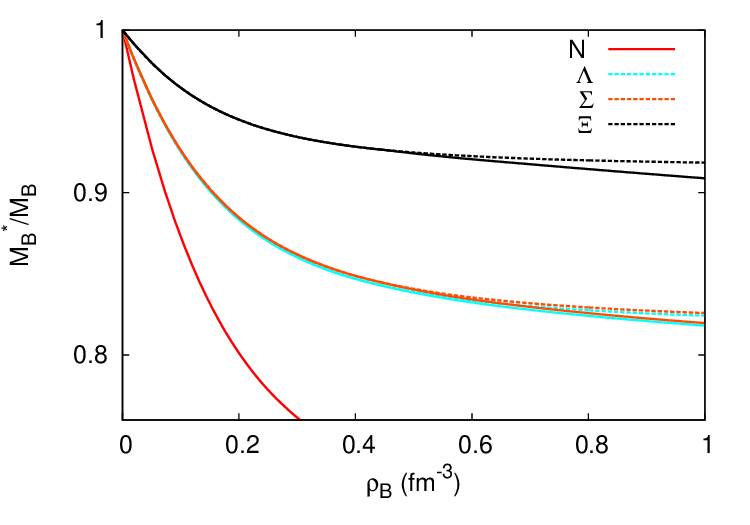}
\caption{Effective mass of baryons. The dotted lines indicate matter 
without the $\sigma^*$ interaction while the continuous lines represent 
baryon mass with $\sigma^*$ interaction.}
\label{fig2}
\end{figure}
%%%%%%%%%%%%%%%%%%%
%%%%%%%%%%%%%%%%%%%%%
%\subsection{Determination of coupling constants}
The quark meson couplings $g_\sigma^q$, $g_\omega=3g_\omega^q$, 
and $g_\rho=g_\rho^q$ are fitted self-consistently for the nucleons to obtain 
the correct saturation properties of nuclear matter binding energy, 
$E_{B.E.}\equiv B_0={\cal E}/\rho_B-M_N=-15.7$ MeV, pressure, $P=0$, and 
symmetry energy $J=32.0$ MeV at 
$\rho_B=\rho_0=0.15$ fm$^{-3}$. 

We have taken the standard values for the 
meson masses; namely, $m_\sigma=550$ MeV, $m_\omega=783$ MeV, 
$m_\rho=763$ MeV, $m_{\sigma^*}=980$ MeV and $m_\phi=1020$ MeV. 
The values of the quark meson couplings, 
$g_\sigma^q$, $g_\omega$, and $g_\rho$ at quark masses $200$ MeV 
are given in Table \ref{table2}.
The value of the $\omega$-$\rho$ coupling term $\Lambda_\nu$, which affects the 
$g_\rho$ coupling \cite{hss2}, is fixed at $\Lambda_\nu=0.05$. In fact, such a 
non-linear $\omega$-$\rho$ term gives rise to effective masses for the 
$\omega$ and $\rho$ mesons, thus softening the vector fields at large 
densities \cite{rabhi}.
\begin{table}[ht]
\centering
\renewcommand{\arraystretch}{1.4}
\setlength\tabcolsep{3pt}
\begin{tabular}{cccccc}
\hline
\hline
$m_q$ & $g^q_\sigma$ & $g_\omega$ & $g_\rho$ & $M_N^*/M_N$ & K\\
(MeV)& & & & & (MeV)\\
\hline
200   &4.36839  &7.40592 &9.39956 &0.83 &242.41\\
\hline
\hline
\end{tabular}
\caption{\label{table2} The quark meson couplings $g_\sigma^q$, 
$g_\omega$ and $g_\rho$ for nuclear matter at quark mass $m_q=200$ MeV. 
$g_\rho$ is determined keeping non-linear coupling fixed at 
$\Lambda_\nu=0.05$. Also shown are the values of the nucleon effective mass 
and the nuclear matter incompressibility $K$.}
\end{table}
 
As discussed in the previous section, we use three types of parameter sets for 
the non-strange and strange meson couplings to the hyperons. For Set I we 
use the SU(6) spin-flavor symmetry. For Set II we follow a mixed scheme where 
the hyperon couplings to the $\omega$-meson  
are fixed by determining $x_{{\omega}B}$. 
The value of $x_{{\omega}B}$ is obtained from the hyperon potentials in nuclear 
matter, $U_B=-(M_B-M_B^*)+x_{{\omega}B}g_{\omega}\omega_0$ for $B={\Lambda},
{\Sigma}$ and $\Xi$ as $-28$ MeV, $30$ MeV and $-18$ MeV respectively. 
For the quark mass $200$ MeV the corresponding values 
for $x_{{\omega}B}$ are given in Table \ref{table3}. 
The value of $x_{{\rho}B}=1$ is fixed for all baryons in this parameter set. 
\begin{table}[b]
\renewcommand{\arraystretch}{1.4}
\setlength\tabcolsep{3pt}
\begin{tabular}{cccc}
\hline
\hline
$m_q$  & $x_{{\omega}\Lambda}$& $x_{{\omega}\Sigma}$&
$x_{{\omega}\Xi}$\\
(MeV) & $U_\Lambda=-28$ MeV& $U_\Sigma=30$ MeV& $U_\Xi=-18$ MeV\\
\hline
200     &0.81316  & 1.43857 & 0.43399 \\
\hline
\hline
\end{tabular}
\caption{\label{table3}$x_{{\omega}B}$ determined for the parameter Set II 
by fixing the potentials for the hyperons.}
\end{table}
%%%%%%%%%%%%%%%

For the Set III we use the SU(3) flavor symmetry to fix the couplings of the 
$\omega$ and $\phi$ mesons with the hyperons. As indicated from the Nagara 
event \cite{nagara}, which suggests the depth of the potential between two 
$\Lambda$s is about $-5$ MeV, we choose the coupling of the strange meson 
$\sigma^*$ to be weak at $2.0$. In fact, using the parameter Set III 
we determine the value of the potential $U_{\Lambda}^{(\Lambda)}$ 
for $\Lambda$ in $\Lambda$-hyperon matter as, 
\begin{equation}
U_{\Lambda}^{(\Lambda)}=-g_{\sigma\Lambda}\sigma_0^{(\Lambda)}-
g_{\sigma^*\Lambda}\sigma_0^{*(\Lambda)}+g_{\omega\Lambda}\omega_0^{(\Lambda)}
+g_{\phi\Lambda}\phi_0^{(\Lambda)},
\end{equation}
which comes out as $-5.28$ MeV.

\begin{figure}
\includegraphics[width=8.cm,angle=0]{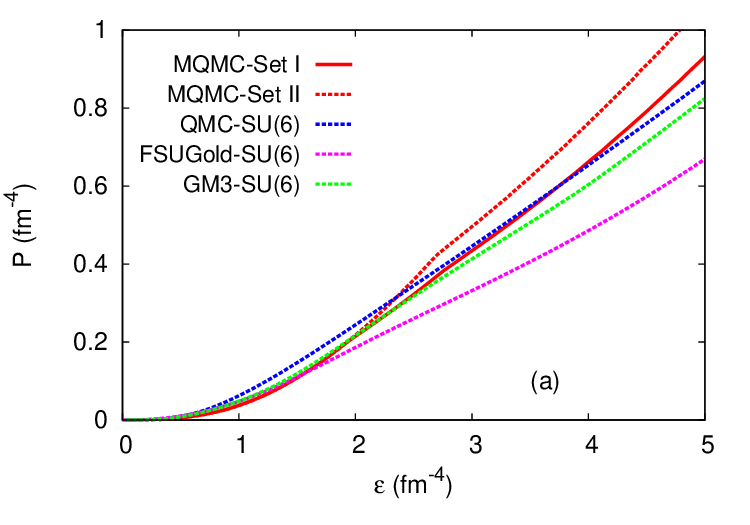}
\includegraphics[width=8.cm,angle=0]{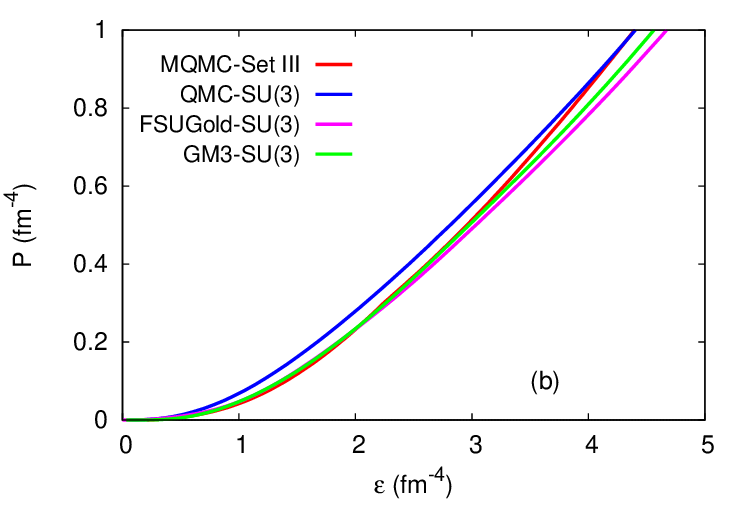}
\caption{Total pressure as a function of the energy density 
at quark mass $m_q=200$ MeV. (a) shows the Set I and Set II of the MQMC 
model as compared to the QMC, FSUGold and GM3 parametrizations in SU(6) while 
(b) shows the comparison of Set III EoS of the MQMC model with 
QMC, FSUGold and GM3 parametrizations in SU(3). 
}
\label{fig4}
\end{figure}
%%%%%%%%%%%%%%%
\begin{figure}
\includegraphics[width=8.cm,angle=0]{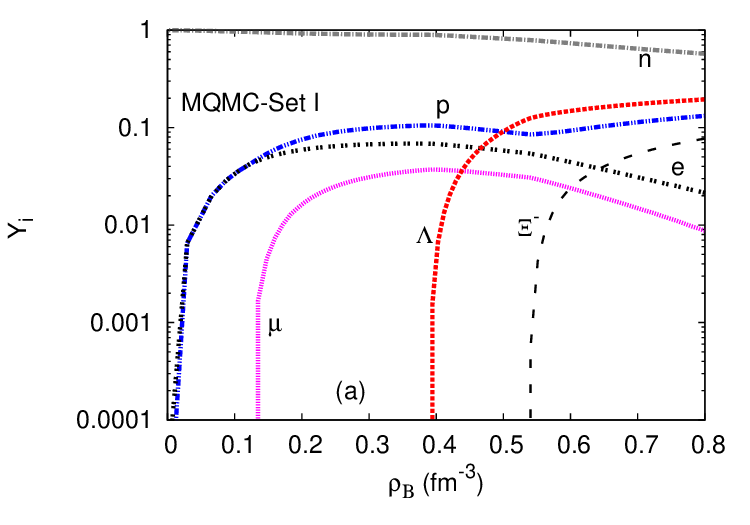}
\includegraphics[width=8.cm,angle=0]{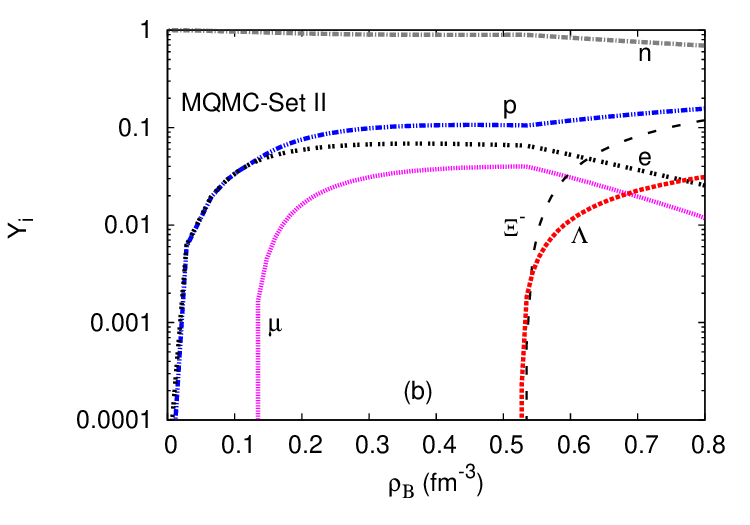}
\includegraphics[width=8.cm,angle=0]{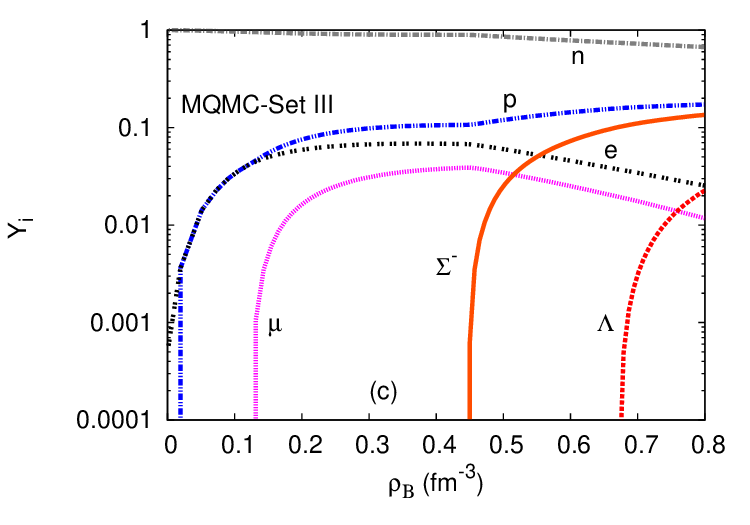}
\caption{Particle fraction for the three parameter sets.}
\label{fig5}
\end{figure}
%%%%%%%%%%%%%%%%%
Fig. \ref{fig2} shows the effective baryon mass, $M_B^*/M_B$, as a function of 
baryon density. At saturation density $\rho_0$ the value of $M_B^*/M_B$ 
is $0.83$ for nucleons.  The effect of the inclusion of strange meson 
$\sigma^*$ on the baryon mass can be observed in the figure. The dotted lines 
indicate the variation in baryon mass in the absence of the $\sigma^*$ meson 
while the continuous lines show the variation with the inclusion of 
$\sigma^*$. The additional strange interaction affects only the hyperons as 
$\sigma^*$ interacts only with strange baryons and decreases the effective 
mass of the hyperons. 

The EoS for the different parameter sets I, II and III in the MQMC model 
is shown in Fig. \ref{fig4}(a) and \ref{fig4}(b) and also compared with the 
results \cite{miyatsu} from QMC \cite{qmc}, FSUGold \cite{fsu} and 
GM3 \cite{gm3} calculations. It is observed that the parameter set II gives 
the stiffest EoS when compared to other SU(6) models. We also observe that 
EoS from SU(3) sets are comparatively more stiffer than the SU(6). This occurs 
due the the enhanced vector-meson couplings to the baryons in SU(3) symmetry. 
  
Fig. \ref{fig5} shows the particle fractions for 
the three types of parameter sets in $\beta$-equilibriated 
matter. Hyperons appear in all the sets choosen, though their threshold 
density of production differs with different meson-hyperon coupling sets. 
$\Sigma^-$ production is preferred in SU(3) coupling Set III indicating 
suppression of $\Xi$ fields at higher densities. Parameter Set II fixes a 
repulsive $\Sigma$-hyperon potential resulting in the absence of 
the $\Sigma$ hyperon the matter distribution.

\begin{figure}[ht]
\includegraphics[width=8.cm,angle=0]{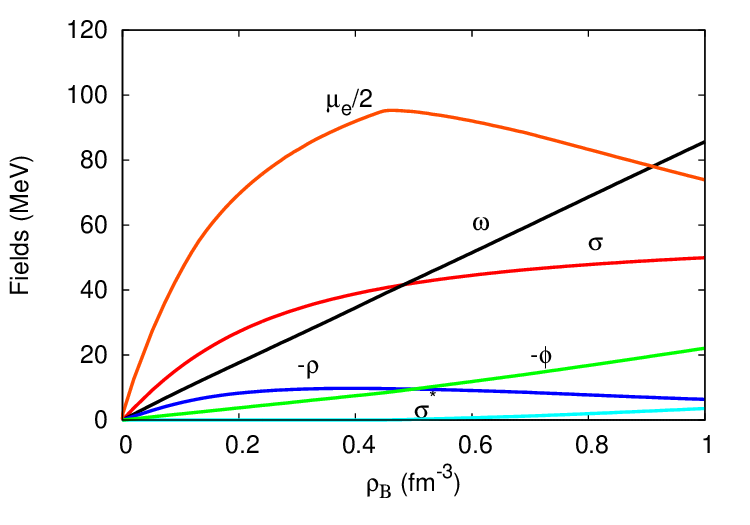}
\caption{Meson fields in the SU(3) set.}
\label{fig6}
\end{figure}

\begin{figure} 
\includegraphics[width=8.cm,angle=0]{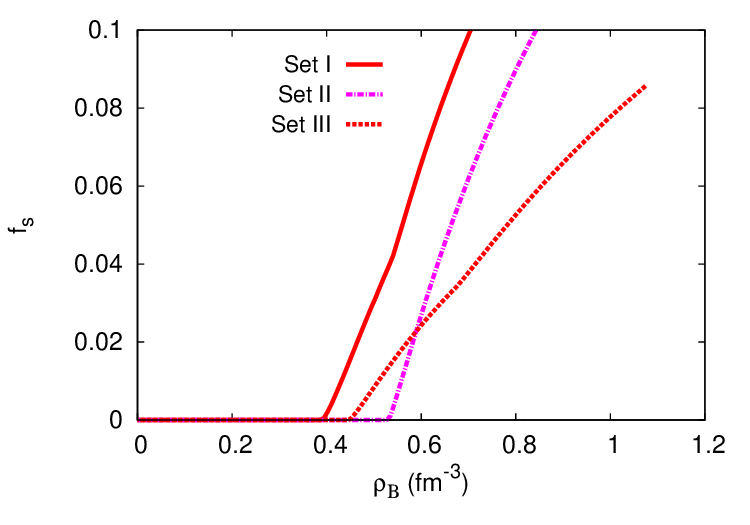}
\caption{Strangeness fraction as
a function of density for Set I, II and III.}
\label{fig6b}
\end{figure}

The meson fields in the SU(3) symmetry are plotted in Fig. \ref{fig6}. It is 
observed that the $\phi$ meson contributes to the baryon 
interactions even at lower densities. This is due to the mixing effect 
considered in SU(3) coupling set. Further more, it is observed that the 
$\sigma^*$ meson appears at densities higher than the density of appearnce of 
the $\Sigma$ hyperon of Set III.
Fig. \ref{fig6b} show the strangeness fraction for the three types of sets 
considered in the  present model. It is observed that the strangeness fraction 
increases at lower densities for the Set I with complete SU(6) couplings. 
Between Set II and Set III, though the strangeness appears earlier in 
Set III, there is sharper increase  in the fraction for Set II indicating 
more strangeness content at lower densities. This is reflected in the EoS 
plot where the EoS for Set II is softer than Set III.

\begin{figure*}
\includegraphics[width=8.cm,angle=0]{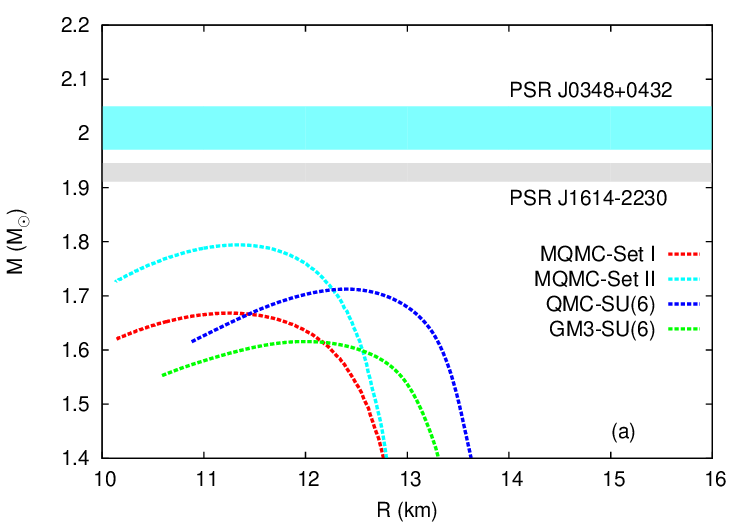}
\includegraphics[width=8.cm,angle=0]{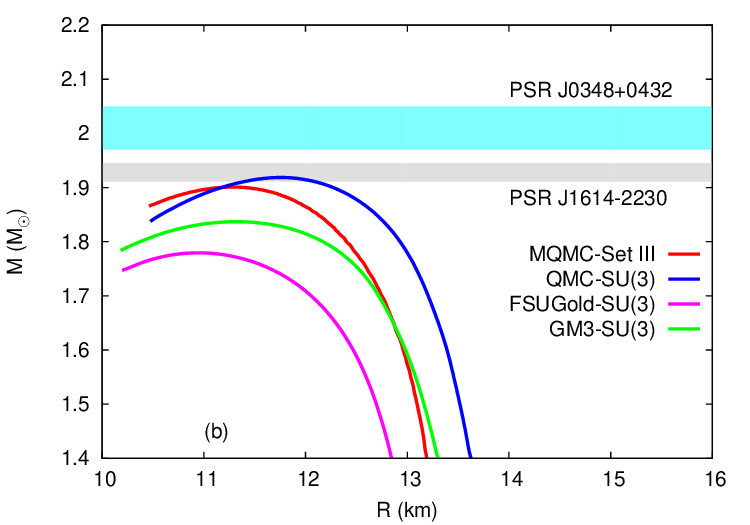}
\caption{Star mass as a function of radius for 
(a) Set I and Set II and (b) Set III at $m_q=200$ MeV along with the 
plots for the QMC, FSUGold and GM3 models.} 
%Also shown is the mass 
%observed for the pulsar PSR J0348+0432 in \cite{antoniadis}.}
\label{fig7}
\end{figure*}

In Fig. \ref{fig7} we plot the mass-radius relations for the different 
parameter sets. It is clearly observed that SU(6) sets fail to achieve the 
maximum mass limit of the pulsar PSR J1614-2230. However, in the SU(3) coupling 
Set III a maximum mass of $1.90$ M$_\odot$ is achieved with a corresponding 
radius of $11.2$ km. The radius corresponding to the canonical 
$1.4$ M$_\odot$ star for set III is $13.1$ km. For the Sets I and II, the 
maximum mass is much lower, but the canonical radius is $12.7$ km. The mass 
and radius for all the three sets are shown in Table \ref{table4}.
\begin{table}[b]
\renewcommand{\arraystretch}{1.4}
\setlength\tabcolsep{3pt}
\begin{tabular}{cccc}
\hline
\hline
$Set$  & M (M$_\odot$) & R (km) & R$_{1.4}$ (km)\\
\hline
Set I     & 1.66 & 11.2 & 12.7\\
Set II    & 1.79 & 11.3 & 12.7\\
Set III   & 1.90 & 11.2 & 13.1\\
\hline
\hline
\end{tabular}
\caption{\label{table4} Mass and radius for different parameter sets. 
Also shown is the radius corresponding to the canonical mass 1.4 M$_\odot$.}
\end{table}

The recent detection of event GW170817 \cite{ligo} and the consequent 
studies involving radius measurements constrain the radii 
of $1.4 M_{\odot}$ mass neutron star, between $9.9<R_{1.4}<13.6$ km. In the 
present work we obtain $R_{1.4}$ between $12.7$ km and $13.1$ km, which is 
within the predicted range. We observe that SU(3) set is favourable for a 
massive star. 

\section{Conclusion}
In the present work we have developed the EoS using a relativistic quark  
model which considers the baryons to be composed of three 
independent relativistic quarks confined by an equal admixture of a 
scalar-vector harmonic potential in a background of scalar and vector mean 
fields. Appropriate corrections to the centre of mass motion, 
pionic and gluonic exchanges within the nucleon are calculated to obtain 
the effective mass of the baryon. 
The baryon-baryon interactions are realised by the quark coupling to the 
$\sigma$, $\omega$ and $\rho$ mesons through a mean field approximation. To 
include the contribution of strangeness, the hyperon-hyperon interaction 
mediated by  $\sigma^*$ and $\phi$ mesons is introduced. 

The strange and non-strange meson couplings to the hyperons are fixed 
using three different techniques based on symmetry considerations and the 
available hyperon-nucleon potentials. The EOS is analyzed for three different 
sets of coupling constants and the effect of such couplings on the 
strangeness fraction is studied. The variations of the maximum mass and radius 
in the present set of parametrizations are determined and it is observed that 
an extension of SU(6) spin-flavor symmetry to SU(3) flavor symmetry is 
favourable to obtain massive yet compact stars. Furthermore, the role of the 
strange $\phi$ meson in sustaining a heavy neutron star is analysed.    
The canonical radius obtained in the present study lies within the 
range of values predicted from studies of the GW170817 event.  

\section*{ACKNOWLEDGMENTS}
The authors would like to acknowledge the financial assistance from 
BRNS, India for the Project No. 2013/37P/66/BRNS. HSS acknowledges the 
the award of CSIR-SRF fellowship.


\begin{thebibliography}{99}
\bibitem{ligo}B. P. Abbott {\it et al.} [LIGO Scientific and 
Virgo Collaboration], Phys. Rev. Lett. {\bf 119}, 161101 (2017).
\bibitem{annala} E. Annala, T. Gorda, A. Kurkela and A. Vuorinen, 
Phys. Rev. Lett. {\bf 120}, 172703 (2018).
\bibitem{janka} A. Bauswein, O. Just, H.-T. Janka, and N. Stergioulas, Astro-
phys. J. Lett. 850, L34 (2017).
\bibitem{margalit} B. Margalit and B. D. Metzger, Astrophys. J. Lett. 850, L19 (2017).
\bibitem{rezolla} L. Rezzolla, E. R. Most, and L. R. Weih, Astrophys. J. Lett. 852, L25 (2018).
\bibitem{shapiro} M. Ruiz, S. L. Shapiro, and A. Tsokaros, Phys. Rev. D 97, 021501 (2018). 
\bibitem{rnm} N. Barik, R. N. Mishra, D. K. Mohanty, P. K. Panda and
T. Frederico, Phys. Rev. C {\bf 88}, 015206 (2013).
%
\bibitem{hss} R. N. Mishra, H. S. Sahoo, P. K. Panda, N. Barik and
T. Frederico, Phys. Rev. {\bf C 92}, 045203 (2015).
%
\bibitem{hss2} R. N. Mishra, H. S. Sahoo, P. K. Panda, N. Barik and
T. Frederico, Phys. Rev. {\bf C 94}, 035805 (2016).
%

\bibitem{barik}N. Barik and B.K. Dash, Phys. Rev. D {\bf 33}, 1925 (1986),
{\it ibid}, Phys. Rev. D {\bf 34}, 2092 (1986).
%
\bibitem{prd} N. Barik and R.N. Mishra, Phys. Rev. D {\bf 61}, 014002 (2000).
%
\bibitem{pdg} J. Beringer et al. (Particle Data Group), Phys.Rev. D 86, 010001 (2012).
%
\bibitem{pais} A. Pais, Rev. Mod. Phys. 38, 215 (1966).
%
\bibitem{miyatsu} T. Miyatsu, M. K. Cheoun, K. Saito Phys. Rev. C 88,
015802 (2013).
%
\bibitem{guichon} P. A. M. Guichon, Phys. Lett. B {\bf 200}, 235 (1988);
P.A.M. Guichon, K. Saito, E. Rodionov, A.W. Thomas, Nucl. Phys. A {\bf 601}
(1996) 349.
%
\bibitem{horowitz01}C. J. Horowitz and J. Piekarewicz, Phys. Rev. Lett. 
{\bf 86}, 5647 (2001).
%
\bibitem{cdover} C. Dover, A. Gal,Prog.Part.Nucl.Phys., 12, 171 (1985).
%
\bibitem{agal} J. Schaffner, C. B. Dover, A. Gal, et al., Annals Phys., 235, 35 (1994).
\bibitem{nijmegen} T. A. Rijken, M. M. Nagels, and Y. Yamamoto, 
Prog. Theor. Phys. Suppl. 185, 14 (2010).
%
\bibitem{tov} J. R. Oppenheimer and G. M. Volkoff, Phys. Rev. {\bf 55},
374 (1939).
%
\bibitem{tov11} R. C. Tolman, Proc. Nat. Acad. Sci. {\bf 20}, 169 (1934).
%
\bibitem{swart} J. J. de Swart, Rev. Mod. Phys. 35, 916 (1963); 37, 326(E) (1965).
\bibitem{rijken} T. A. Rijken, V. G. J. Stoks, and Y. Yamamoto, Phys. Rev. C 59, 21 (1999).
\bibitem{weiss} S. Weissenborn, D. Chatterjee, and J. Schaffner-Bielich, Phys. Rev. C 85, 065802 (2012); Nucl. Phys. A 881, 62 (2012).
%
\bibitem{umesh} U. Garg and G. Col\`o, arXiv:1801.03672.
%
\bibitem{rabhi} C. Providen\^cia and A. Rabhi, Phys. Rev. C {\bf 87}, 
055801 (2013).
%
\bibitem{nagara} H. Takahashi et al., Phys. Rev. Lett. 87, 212502 (2001).
\bibitem{qmc} T. Miyatsu, T. Katayama, and K. Saito, Phys. Lett. B 709, 242
(2012).
\bibitem{fsu} B. G. Todd-Rutel and J. Piekarewicz,Phys. Rev. Lett. 95, 122501
(2005).
\bibitem{gm3} N. K. Glendenning and S. A. Moszkowski, Phys. Rev. Lett. 67,
2414 (1991).
\end{thebibliography}
\end{document}